# TEI Piraeus students' knowledge on the beneficial applications of nuclear physics: Nuclear energy, radioactivity - consequences

Mirofora Pilakouta
Department of Physics Chemistry and Material Technology
T.E.I. of Piraeus, Greece
Tel: 2105381583  E-mail:mpilak@teipir.gr

The recent nuclear accident in Japan revealed the confusion and the inadequate knowledge of the citizens about the issues of nuclear energy, nuclear applications, radioactivity and their consequences
In this work we present the first results of an ongoing study which aims to evaluate the knowledge and the views of Greek undergraduate students on the above issues. A web based survey was conducted and 131 students from TEI Piraeus answered a multiple choice questionnaire with questions of general interest on nuclear energy, nuclear applications, radioactivity and their consequences. The survey showed that students, like the general population, have a series of faulty views on general interest nuclear issues. Furthermore, the first results indicate that our educational system is not so effective as source of information on these issues in comparison to the media and internet

1. **Introduction**

The recent nuclear disaster in Japan has increased the interest and the fears of general population in all over the world about nuclear issues, radiations in general and especially radioactivity, and their consequences in human health and the environment. There is a great number of issues that are related to nuclear physics and its applications and touch, political, economical and social aspects of our lives. So it is very important for the citizens to have an acceptable level of knowledge on these issues.
Several studies have been conducted in different countries [1-3], trying to evaluate the knowledge, perceptions and views of secondary school students about various nuclear issues. Other surveys focus to investigate the attitude of students [4] or general population [5] towards nuclear power and nuclear applications. The general conclusion is that there is poor understanding about these issues and it is underlined that the educational system, especially the secondary school, should give more attention to make the students (the future citizens) aware of nuclear issues [1-4].
The secondary school curricular in Greece contains some topics on nuclear physics but they are not always of first priority either for the teachers or for the majority of the students. Taking into account that secondary education is the last chance for the largest part of general population to achieve reliable information on these issues, we assume that most of the Greek undergraduate (non major in physics) students have more or less the same confusion on nuclear issues as the general population.

In this work we present results from a recent survey conducted at TEI Piraeus that evaluates the knowledge and views of TEI Piraeus students on some nuclear issues of general interest.

This is the first step of an ongoing research which aims
- To investigate the knowledge of Greek students on some nuclear application issues,
- To compare student's knowledge to that of general population
- To propose ways to link the educational material with topics of current interest, in order to motivate students to explore these concepts in a more serious manner.

Using the advantage of the increased interest (due to Fukushima accident) on these issues, we used the survey as an extra tool to motivate students to participate in a seminar about the nuclear energy and the nuclear accidents. The seminar aimed to make the students aware of both, the useful applications of nuclear physics in our life and the harmful effects of radioactivity in health and environment.

## 2. Survey development and administration

Since our target population is students (engineers) that have a positive attitude towards internet, the survey was conducted using an online questionnaire. The questionnaire was created using a form in Google Docs and was posted for a week on the Internet accompanied by the announcement of the seminar (about Nuclear energy, accidents, radioactivity and consequences) that would take place in a week. An invitation for filling the questionnaire was posted on the forum of the automation department students. The questionnaire was also sent in a number of students by e-mail and to Secretariats of the STEF faculty of TEI Piraeus. Finally, students (from civil and mechanical engineering departments) who were exercising in the physics lab that week were also asked to fill optionally the questionnaire using a computer in the laboratory.

The questionnaire was filled by 131 students. Data were automatically collected in a spreadsheet (in Google Docs) and were analyzed after the completion of the survey.

A great number of questions could have been incorporated in this questionnaire. In this preliminary study the questionnaire was consisted of 10 multiple-choice questions. Eight of the questions were examining basic and attractive topics related to nuclear issues, based on commonly documented misconceptions [1-3]. The other two questions ask for the source of their information and their attitude towards the seminars of general interest in the TEI. The questions are given in Table 1.

| QUESTIONNAIRE ||
|---|---|
| QUESTION | OPTIONS |
| **Q1: Which of the following radiations may produce genetic problems** | a) Visible radiation <br> b) Ultraviolet radiation <br> c) Gamma rays <br> d) Mobile phone radiation <br> e) Don't Know |
| **Q2: The gamma rays are electromagnetic waves emmited by** | a) nuclei and have higher energy than X-rays <br> b) atoms and have lower energy than X-rays <br> c) Laser devices <br> d) Don't Know |
| **Q3: The ionizing radiation which mainly contributes to the total irradiation of humans during their lives, comes from:** | a) Nuclear Plants <br> b) Nuclear accidents <br> c) Medical tests <br> d) Natural Radioactivity <br> e) Don't Know |
| **Q4: How is energy produced in a Nuclear plant?** | a) Spontaneous disintegration of heavy nuclei <br> b) The fusion nuclear reactions <br> c) The fission nuclear reactions <br> d) Don't Know |
| **Q5: Which European country covers more than 70% of its energy needs using nuclear energy?** | a) Belarus <br> b) France <br> c) Germany <br> d) Great Britain <br> e) Don't Know |
| **Q6: Which is the pair of radioactive elements of high concern that are released in the environment during a nuclear accident** | a) Iodine and Uranium <br> b) Plutonium and Uranium <br> c) Cesium and Iodine <br> d) Don't Know |
| **Q7: How many deaths are directly attributed to the nuclear accident during the first three months after Chernobyl disaster in 1986?** | a) 5-50 <br> b) 50-100 <br> c) More than 1000 <br> d) More than 100000 <br> e) Don't Know |
| **Q8: Which is the most unknown application of ionizing radiation for you?** | a) Medicine (diagnosis, treatment) <br> b) Sterilization of instruments used in medicine or food products <br> c) Research-Industry applications (radiography) <br> d) I Know all of them <br> e) I don't know any of them |

**Table 1:** Questions on general interest Nuclear Issues

### 3. Survey outcomes

In the following we discuss the result for every individual question and try to analyze the underlying reasons for some incorrect answers.

- **Which of the following radiations may produce genetic problems?**

A majority, 89 percent of the students, gave correct answer in this question (gamma rays), while 11 % seems not to understand the difference between ionizing and non ionizing radiation, which is a common misunderstanding for general population as well.

- **Which is the origin and nature of gamma rays?**

The majority of the students (71%) seem to know the origin and distinguish γ radiation from X-rays. However, a total of 21 % demonstrated confusion about the nature of gamma rays, when another 8% admit that they don't know. It is notable that 8% of the students have the faulty conception that gamma rays can be produced by laser devices. This is probably related to the confusion about the term "radiation" which is used to describe several sources of radiation.

- **Where does the ionizing radiation, which mainly contributes to the total irradiation of humans during their lives, come from?**

Only one out of three (34%) of the students knows that the main contribution to human irradiation comes from natural radioactivity. The majority, 40%, believe that the main source of irradiation comes from medical examinations, 15% from nuclear plants or nuclear accidents and a minority 11% answers that they do not know.
The belief that the contribution of medical examinations is higher than the contribution of other sources is related to the general fear about medical examinations using ionizing radiation as well as to the fact that general population have poor knowledge about natural radioactivity.

- **How is energy produced in a nuclear plant?**

Two out of three of the students (66%) know that energy is released due to the nuclear fission. Far fewer answers concerned to the nuclear fusion (13%), or spontaneous disintegration of heavy nucleus (7%). The percentage of the students that declared they don't know was 14%.

- **Which European Country covers more than 70% of its energy needs using Nuclear Energy?**

Most of the students did not know this information. Only 26 % of the student's knew that France is pioneer in using nuclear energy. Another 27% declared they don't know and 2% thinks is the Great Britain. It is noticeable that 21% of the students considered that Belarus is the country that mostly depends on nuclear energy, although this country does not have nuclear plants yet. This wrong impression may be attributed, firstly to the fact that this country was mostly affected by the nuclear accident at Chernobyl in 1986 and secondly to the poor understanding of the global effects of a nuclear accident.

- **Which is the pair of radioactive elements of high concern that are released in the environment during a nuclear accident?**

A big number of students (58%) have the wrong perception that Plutonium and Uranium are the radioactive elements that are mainly released in the environment.
Only 18% answered correctly (cesium and iodine). A minority of 12% think of iodine and plutonium and 11% declared that they don't know. This indicates that the majority of the student confuses the nuclear fuel with the reaction products.

- **How many deaths are directly attributed to the nuclear accident during the first three months after Chernobyl disaster in 1986?**

Only 4% of the students knew that less than 50 deaths can be attributed directly to the radioactive contamination due to the Chernobyl accident. The majority of 96% either declare that they don't know (28%), or believe that deaths were more than 10.000 (35%), more than 100.000 (27%) or about 50-100 (6%). The above result shows the huge misconception, most of the students (and general population) have about the direct and the aftermath effects of radioactive contamination. The main reason for this faulty impression is that the media and several organizations (UN, Atomic Energy Agency, WHO) report mainly the potential effects of the ionizing radiation. Furthermore estimations of the number of deaths potentially resulting from Chernobyl accident, vary enormously between experts [2, 6].

- **Which is the most unknown application of ionizing radiation for you?**

The responses indicate that only 31% of the students know all the proposed useful applications of ionizing radiation. The most unknown application (24% stated that is the most unknown for them) was the sterilization of instruments used in medicine or of food products. It is notable that 15% of the students stated that they ignore the use of ionizing radiation in medicine (diagnosis, treatment) although in our country it is used in excessive degree. Finally 16% consider the Research-Industry applications as the most unknown for them and 14% didn't know any of them.

- **What is your main source of information about radioactivity issues?**

Overall, an average of three out of four students (73%) has information from mass media (internet (56%) and radio-TV (17%)). A small number of students (13%) have other source of information and only 14% indicate school as the source of their information about radioactivity topics. This indicates that our school fails to offer the appropriate knowledge about these serious issues that affect our lives.

- **Students' appreciation of the seminars related to issues of general interest**

The majority of the students (98%), believe that seminars on issues of general interest (like the above mentioned) should be organized in TEI (128 out of 131 students gave positive answer). This indicates a positive attitude about non formal ways of learning.

In figure 1, the percentage of the correct, incorrect, and "not know" answers in the first 7 of the questions are summarized. In this graph we can see that in more than half of the questions, the percentage of students that give incorrect answers is higher than those who answered correctly and much higher than that of the students that declare that they don't know. This suggests that our students have inadequate and confused knowledge in these topics.

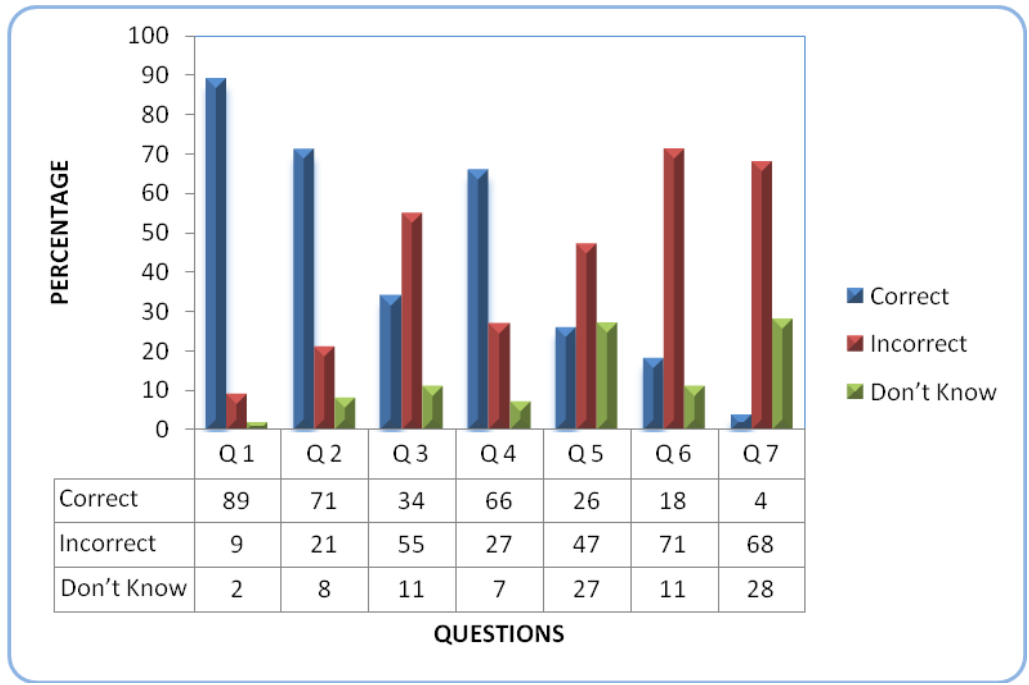

**Figure 1:** Percentage of the correct, incorrect, and "not know" answers in some of the questions

### 4. Discussion and conclusions

The first results of an ongoing research which aims to investigate the knowledge of Greek undergraduate students on some nuclear application issues were presented.

The above results show that the TEI Piraeus students have in general low awareness of the above nuclear issues although they seem to have increased interest about these issues. The students show a highly positive attitude about nontraditional types of education, like general interest seminars, and this is something that we, the teachers, must use in order to improve our students knowledge about modern physics issues. Furthermore, a serious effort must be made in secondary school to motivate students' interest for both, the useful applications of nuclear physics in our life and the harmful effects of radioactivity in health and environment.

The outcome of this survey, in combination with some students' comments, will be used to develop a revised survey that will be administered to a greater number of Greek students. In the mean time, a similar survey, with slight revision of some questions, is to be conducted soon with target group the administrative staff of TEI Piraeus, to compare the "general population" knowledge and views in nuclear issues with that of the "non major in physics" students. The identification of the most common misconceptions and faulty views of students and general population will contribute in the development of educational material that will improve the understanding of nuclear issues related to our everyday life.

## Acknowledgements

The author would like to thank the students of the Automation Department Dimitris Pantelis and Enkeleda Bocaj for their assistance in the conduct of this survey.